\documentstyle[preprint,aps,eqsecnum]{revtex}
 
\def\beq#1{\begin{equation} \label{#1}}
\def\eeq{\end{equation}}
\def\bra#1{\left\langle #1\right\vert}
\def\ket#1{\left\vert #1\right\rangle}

\begin{document}
{
\tighten

\title{Physics of Debye-Waller Factors}
 
\author{ Harry J. Lipkin\,$^{a,b}$}
 
\address{ \vbox{\vskip 0.truecm}
  $^a\;$Department of Particle Physics \\
  Weizmann Institute of Science, Rehovot 76100, Israel \\
\vbox{\vskip 0.truecm}
$^b\;$School of Physics and Astronomy \\
Raymond and Beverly Sackler Faculty of Exact Sciences \\
Tel Aviv University, Tel Aviv, Israel}
 
\maketitle  \begin{abstract}    This note has no new results and is therefore
not intended to be submitted to a "research" journal in the foreseeable future,
but to be available to the numerous individuals who are interested in this
issue. The Debye-Waller factor is the ratio of the coherent scattering or
absorption  cross section of a photon or electron by particles bound in a
complex system to the  value for the same process on an analgous free particle.
It is often interpreted also as the probability of the coherent process,
normalized to unity, with the difference between unity and the Debye Waller
factor interpreted as the probability of incoherent processes. The Debye-Waller
factor is then interpreted as a measure of decoherence. The breakdown of this
description for a test particle which cannot  give or lose energy is not
generally appreciated. Prime examples are: Bragg scattering, the M\"ossbauer
effect and related phenomena at zero temperature. The physics of the change in
the interpretation of the Debye-Waller factor is summarized here in a hopefully
pedagogical manner.

\end{abstract}

} 

\section{Inroduction}

When a particle is scattered by another particle which is bound in a complex
system, the probability of elastic scattering with no change in the state of
the complex system is proportional to a quantity called the Debye-Waller 
factor. This originally arose in the scattering of X-rays by atoms in a 
crystal, where the probability of coherent scattering was proportional to the
Debye-Waller factor, which was always less than unity, and depended upon the
temperature of the crystal. There was also a probability of incoherent 
scattering. The sum of the squares of the incoherent and coherent scattering 
amplitudes from a given atom was always equal to the square of the scattering
amplitude from a free atom. The Debye-Waller factor might be interpreted as
a measure of decoherence. However, this is not generally true, and the physics
of decoherence and dephasing in processes where a test particle cannot gain or
lose energy has been
discussed in detail\cite{Imry}. We focus here on the probability interpretation
and the relation between analogous processes on free vs. bound particles.

In any process described by the Golden Rule of time-dependent perturbation
theory that involves a final state with at least one particle in a continuum
the transition probability per unit time is proportional to the product of the
square of the  transition matrix element and the density of allowed final
states. In X-ray scattering by crystals and in M\"ossbauer transitions the
energy of the outgoing X-ray is so large compared to lattice excitation
energies that the difference  in phase space over the entire elastic and
inelastic spectrum can be negligible. In other cases it may not be negligible.
In all cases the square of the  transition matrix for coherent elastic
scattering is reduced by the Debye-Waller factor from the square of the matrix
element for scattering from a free particle. But this reduction is not a result
of decoherence. 

This can be seen by simply writing the expression for the transition matrix 
element between a given initial state $ \ket {i} $ of the complex system 
and a given final state  $ \ket {f} $ for the scattering of an incident
particle by a component of the complex system whose co-ordinate is $\vec x$ 
\beq{QQ1}
\bra{f} T \ket {i} = g(\vec k) \bra{f} e^{i \vec k \cdot \vec x}  \ket {i} 
\eeq
where $\vec k$ is the momentum transfer and $g(\vec k)$ is the strength of the
interaction causing the scattering. 

This expression clearly satisfies the sum rule,
\beq{QQ2a}
\sum_f |\bra{f} T \ket {i}|^2 = |g(\vec k) |^2   
\eeq 
while the square of the transition matrix for scattering without changing the 
state of the complex system is given by     
\beq{QQ2b}
|\bra{i} T \ket {i}|^2 = 
|g(\vec k) |^2 |\bra{i} e^{i \vec k \cdot \vec x}  \ket {i}|^2    
\eeq 

For the case where the initial state is a free particle, there is only a single
final state in the sum and the quantity $|g(\vec k) |^2$ is just the square of
the transition matrix element for scattering on a free particle. This naturally
leads to the  description in which the square of the transition matrix element 
for scattering without changing the state of the complex system is given by the
product of the quantity  $|g(\vec k) |^2$ and the Debye-Waller factor  
$|\bra{i}  e^{i \vec k\cdot \vec x} \ket {i}|^2$.  For cases where the
transition probability  is proportional to the square of the transition matrix
element with the same proportionality factor for all final states $ \ket {f} $
the  Debye-Waller factor gives the relative probability of elastic
scattering vs. inelastic scattering, where the total probability of elastic and
inelastic  scattering is normalized to unity. 
\beq{QQ2c} \sum_f |\bra{f} e^{i
\vec k \cdot \vec x}  \ket {i}|^2  = 1 
\eeq  
This picture of the Debye-Waller
factor as defining a relative probability of coherent vs. incoherent
scattering, or of elastic vs. inelastic processes has entered the folklore in
many areas of physics, but it is not strictly correct. It is an approximation
which holds only in a certain kinematic region. 

A problem arises because the transition probability is not simply proportional
to the square of the transition matrix element. There is also an additional
kinematic phase space factor. In the case of X-ray scattering by crystals and
the M\"ossbauer effect, the change in the phase space factor over the elastic
and inelastic spectra is negligible and one can accept the probability
interpretation. When the available phase space for the final state is not a
constant over the entire spectrum, the expression for the elastic transition,
(\ref{QQ2b}) is still valid, but the Debye-Waller factor can no longer be
interpreted as a probability. It represents the reduction factor for the
elastic transition from a bound system relative to that for the scattering from
a free particle, but the sum of the elastic and inelastic transitions no longer
satisfies the sum rule (\ref{QQ2a}) and there is no longer a normalized
probability. 

Consider, for example, the scattering of a photon by an Einstein crystal in its
ground state;  i.e. at zero temperature. If the energy of the photon is less
than the energy required to excite one phonon, there can only be elastic
scattering. The transition matrix elements for transitions to all states of the
lattice are still given by eq. (\ref{QQ1}). and  the cross section
for elastic scattering by an atom in the lattice is still reduced by the
Debye-Waller factor from the value of the cross section for elastic scatering
by a free atom. But there is no inelastic scattering and the total scattering
cross section is reduced frrom the free atom value. The sum rule (\ref{QQ2a})
relates  transition matrix elements but not probabilities. 

If the scattering is observed at a Bragg angle, the coherence of the
amplitudes scattered from different atoms in the crystal gives the normal Bragg
peak in the angular distribution with an intensity for coherent scattering
proportional as usual to the Debye-Waller factor. But here there is no
incoherent scattering. All the scattering is coherent. The Debye-Waller factor 
does not produce decoherence. At a finite temperature where there is a
possibility that the photon can gain energy in the scattering process, this
inelastic scattering is incoherent. But the variation with temperature of the
coherent scattering is described by the Debye-Waller factor independently of
the additional incoherent inelastic scattering. The Debye=Waller factor does
not produce incoherence nor decoherence.

When this is applied to the scattering of an electron at the top of a fermi 
sea at zero temperature by either an electron gas in a metal or by impurities,
each individual scattering is reduced by its individual Debye-Waller factor.
But there is no inelastic scattering because the electron can neither gain nor
lose energy. It cannot gain energy from the rest of the system at zero
temperature and it cannot lose energy since it is at the top of a Fermi
sea. The total  scattering cross section is reduced. At each interaction the
probability that  the electron escapes without scattering is increased above
the normal  probability for the case where it has the same energy and the fermi
sea is  unoccupied. Thus mean free paths become larger and other effects
normally  ignored become more important.

\section{Debye-Waller factors for macroscopic systems}

In a completely different context we note that the Debye-Waller factor plays an
interesting role in interference experiments with photons scattered by mirrors.
One may be  tempted to consider the mirrors as classical objects, and use
classical  kinematics to describe the momentum and energy transfer to the
mirror when the photon is scattered. Since this momentum and energy is tiny,
one can envision a very sensitive device which can use this tiny energy as a 
trigger (e.g. by exploding a bomb) to give evidence of the scattering of the
photon by this particular mirror and destroy the interference. Conversesly, if
the photon is detected without exploding the bomb this can be evidence that the
photon was scattered by another mirror that was not coupled to the bomb.

The fallacy in such arguments is that  the mirrors  themselves are subject to
the laws of quantum mechanics. Since they must be bound to a fixed position in
space, they will be described by quantum states with a discrete energy
spectrum.  The allowed values of energy transfer to the mirror when the photon
is  scattered must correspond to the energy differences bvetween energy levels
of the bound mirror.  The scattering of a photon by a mirror will therefore be
either elastic, in which case the mirror remains in the same initial energy
level and there is no energy transfer to the mirror, or it will be inelastic,
in which case the mirror jumps to a different energy level. 

The tiny energy transfer given by the classical result for the scattering of a
photon by a mirror can be interpreted  by the correspondence principle as only
the $average$ energy transfer. In all practical cases this tiny energy transfer
is very much smaller than the energy level spacings. Thus the probability of
elasic scattering in which there is no energy transfer and therefore no
probability of triggering a bomb will be very close to unity. It is given by
the Debye-Waller factor   $|\bra{i}  e^{i \vec k \cdot \vec x} \ket {i}|^2
\approx e^{-k^2<x^2>}$ where here $\ket {i}$ denotes the initial quantum state
of the mirror,  $k$ is the momentum transfer and $<x^2>$ is the mean square 
fluctuation in the position of the mirror. Since the position fluctuation must
be tiny in comparison with the wave length of the photon in any realistic
experiment, the Debye-Waller factor will be very close to unity, and the
probability that sufficient energy is transfered to the mirror to trigger a
bomb is vanishingly small.

Another way to describe the same physics is to note that the kinetic energy
transfer to the mirror by the recoil momentum of scattering a photon is not
only tiny compared with the  spacing between the energy levels of the bound
states of the mirror, but also tiny in comparison  with the zero-point kinetic
energy of the mirror in its bound state. Since the only allowed energy 
transfer is the energy difference between energy levels, the probability that a
tiny  change in kinetic energy can produce a quantum jump to another energy
level is negigible.

Another context with similar physics is in the detection of neutrinos in a
neutrino oscillataion experiment.  Neutrino  oscillations arise because the 
neutrino state incident on the neutrino detector is a well-defined coherent
mixture of states having different masses, different momenta and different
energies. Particle physicists tend to overlook the fact that the neutrino
detector is a quantum-mechanical condensed matter system and that the neutrino
detection process involves a Debye-Waller factor. Coherence is preserved
between incident neutrino states having the same energy, different masses and
different momenta because the Debye-Waller factor is very close to unity and
the transition from these  different mass eigenstates lead to the same final
state of the detector. However, incident neutrinos with different energies
produce final states of the detector with different energies and all phase
information between neutrino  states with different energy is lost. This causes
confusion among particle physicists who are accustomed to treat energy and
momentum as different components of the same four-vector and cannot understand
why states with the same energy should be coherent and states with the same
momentum and different energies are not. But the interaction with the detector
breaks the energy-momentum symmetry. The detector is a condensed matter system
initially in thermal equilibrium and described in its rest system  by a density
matrix which is diagonal in energy but not in momentum.


\begin{references}
 \bibitem{Imry} Yoseph Imry, cond-mat/0202044
 \bibitem{Mossb2}
{H.J. Lipkin,
Phys.Rev. A42 (1990) 49}
 \bibitem{Mossb}
{H.J. Lipkin,
Hyperfine Interactions, 72 (1992) 3.}
\end{references}
\end{document}